\documentstyle[12pt,epsf,amstex]{article}
\begin{document}
\baselineskip 24pt plus 2pt

\newcommand{\beq}{\begin{equation}}
\newcommand{\beqa}{\begin{eqnarray}}
\newcommand{\eeqa}{\end{eqnarray}}
\newcommand{\dida}[1]{/ \!\!\! #1}
\renewcommand{\Im}{\mbox{\sl{Im}}}
\renewcommand{\Re}{\mbox{\sl{Re}}}
\def\simge{\hspace*{0.2em}\raisebox{0.5ex}{$>$}
     \hspace{-0.8em}\raisebox{-0.3em}{$\sim$}\hspace*{0.2em}}
\def\simle{\hspace*{0.2em}\raisebox{0.5ex}{$<$}
     \hspace{-0.8em}\raisebox{-0.3em}{$\sim$}\hspace*{0.2em}}

\begin{titlepage}


\hfill{TRI-PP-99-03}

\vspace{1.0cm}

\begin{center}
{\large {\bf A Study of the Charged Scalar in the Zee Model}}

\vspace{1.2cm}

G. C. McLaughlin and
J. N. Ng \\

\vspace{0.8cm}

TRIUMF, 4004 Wesbrook Mall, Vancouver, BC, Canada V6T 2A3\\[0.4cm]
\end{center}

\vspace{1cm}

\begin{abstract}
An extension of the Zee model involving a light right handed neutrino, 
$\nu_{R}$ is considered.
We update constraints on couplings between the bilepton scalar, 
the active neutrinos, $\nu_{R}$ and the charged leptons.  
We find that the most stringent constraint currently comes from 
measurements limiting the width of the decay $\mu \rightarrow e \, \gamma$.
These are used to predict the upper bound on violation of lepton universality
in leptonic W boson decays and rare Z decays, such as ${\rm Z} \rightarrow
e \mu$.  \\[0.3cm]

\end{abstract}

\vspace{2cm}
\vfill
\end{titlepage}

Evidence \cite{Con} is now mounting that points towards neutrinos having 
small but
finite masses.  The neutrino mixing explanations which have been invoked to
explain the deficit of solar neutrinos, 
the atmospheric neutrino anomaly and the 
LSND results, require neutrino mass squared differences in the range, 
$\delta m^2$  
of $10^{-6}$ to a few  $\rm {eV}^{2}$.  The exact values depend on the
details of the models used to analyze the data.  The standard explanations for
each phenomenon involve mixing between two flavors of neutrino.  For
example the atmospheric neutrino anomaly  is considered to be 
 large angle mixing between the muon neutrino and something else,
which is not the electron neutrino.  These three standard interpretations
are not easily reconciled without four neutrino states.  In an alternative, 
three neutrino mixing, it 
is difficult to accommodate the zenith angle dependence
of the SuperKamiokande atmospheric results \cite{Naka}.   

It is
generally accepted that neutrinos with masses in the above mentioned
range necessitate the extension of the Standard model (SM) of electroweak
interactions. Early simple
models involving neutrino mass and/or a fourth neutrino state  
were constructed in \cite{Jarl} and \cite{Chang}.
The simplest way to generate neutrino mass is to add at least one SM singlet right-
handed fermion, denoted by $\nu_R$, to the matter content. Indeed Grand 
unified models such as SO(10) naturally contain such a neutral fermion.
In such a model, 
masses for the ordinary neutrinos can be elegantly explained by
the seesaw mechanism. For masses as small as those predicted by
solar, atmospheric and LSND neutrino mixing, 
the SM singlet, $\nu_R$ is required  to have
a mass in the range of $10^{10}-10^{12}$ GeV.    
The seesaw mechanism does not easily accommodate 
a light fourth neutrino.  Therefore, in accepting such a scheme one also 
implicitly accepts the view that one or more of the explanations of the evidence for
neutrino mixing is misleading. Furthermore, there are several uncertainties which arise 
when utilizing the seesaw mechanism to
produce light neutrinos. The means of 
 obtaining this high energy SM singlet is highly
model dependent. The energy scale of the heavy $\nu_R$ is arbitrary and 
reproducing  any of the mass and mixing schemes usually requires model
embellishments.   

In view of these uncertainties
it is important to investigate scenarios or  models which relate
the small neutrino mass with new physics at the weak scale or just above,
i.e. $< O(\mathrm TeV)$. This alternative has the phenomenological advantage that 
such models may be easily constrained or perhaps tested. The simplest
scenario was constructed by Zee\cite{Zee} many years ago. In the original
formulation only the ordinary
left-handed neutrinos are employed and their masses are generated by 
radiative corrections.  Since the masses produced in this way are naturally small,
it is unnecessary to invoke a large mass scale. 
A light $\nu_R$ can also be incorporated as an extension to the model.
However, this is done at the price of giving up the predictability of the 
neutrino masses
and mixings \cite{Wolf,Smir,Jarl2}. To date the bulk of the literature on the
Zee model is devoted to the study of the neutrino mass matrix and neutrino
mixings. The charged scalar meson, $S^-$, which we shall refer to as the
bilepton scalar is the crucial agent that carries
the lepton flavor violation (LFV) necessary for neutrino mass generation and has been
relegated to a secondary role in all studies thus far.

In this paper we study the Zee model as the simplest model of
lepton number violation at the weak scale.  In view of all
of the data on neutrino oscillations, we have  augmented it with a 
$\nu_R$ \cite{fugu}.
 We focus on the physics involved with virtual effects 
of  the charged scalar. Since no charged scalar has been found up to 
LEPII energies,  
we assume its mass to be greater than 100 GeV. 
Due to the simplicity of this model, the number
of free parameters is relatively few.  Furthermore, these can be
 tightly constrained by current
precision measurements. We begin by updating all the constraints 
stemming from muon decays  and tau decays. 
We also discuss the impact of the next
generation of Michel parameters measurements \cite{E614} on the model.
We find that the strongest constraint currently comes from 
$\mu \to e\gamma$ decay.
We then present new results for leptonic universality tests involving
physical W boson decays. This is particularly important in view of the 
large sample of W bosons which will be obtainable from the LHC and the next
linear collider (NLC). We also calculate corrections to the left-handed
charged lepton and Z boson couplings as well as the right-handed charged lepton and Z couplings.
As we shall see below these two types of corrections take very different
forms. We also give the predicted widths
of the rare Z decays such as Z $\to e\mu$. As far as we can determine
these results have not been presented before.

       Without further ado the following Lagrangian is added to the
SM: 
\begin{equation}
\begin{split}
{\cal{L}}_{\mathrm S} =& [ f_{12}(\overline{\nu_e^c}\mu_L 
-\overline{\nu_\mu^c}e_L) +
f_{13}(\overline{\nu_e^c}\tau_L -\overline{\nu_\tau^c}e_L) 
+ f_{23}( \overline{\nu_\mu^c}\tau_L - \overline{
\nu_\tau^c}\mu_L)  \\ 
&   + g_1\overline{\nu_R^c}e_R + g_2\overline{\nu_R^c}\mu_R + 
g_3\overline{\nu_R^c}\tau_R ]S^+
+h.c.\;,
\end{split}
\end{equation}
where $S^+$ is the Zee scalar boson with hypercharge Y=2. The $f_{ij}$ and
$g_i$ are Yukawa couplings and make up six free coupling parameters of 
the model. If the
all the g's are set to zero one recovers the original Zee model. The 
SU(2)$\times$U(1) charges and the lepton number L of the leptons 
and $S^-$ are presented
in Table I. It is seen that the Lagrangian conserves total lepton number;
 however, individual e, mu or tau number is violated. We have not included
the scalar potential for the $S^-$ and its interaction with the SM Higgs 
doublet as they are not needed here. It suffices to note that the Zee boson
can develop a mass either by spontaneous symmetry breaking via coupling
to the SM Higgs doublet and/or explicitly through a bare mass term. In
either case the physical mass is another free parameter which we
denote by $M_{S}$. The above Lagrangian has been used to study neutrino
oscillations in 
\cite{Smir,fugu}. (It was found that the $\nu_R$ significantly changes the
phenomenology of neutrino oscillations from that of the three light neutrinos
case and a fit to all data can be achieved). It will be seen below that since 
we have assumed a light mass for this particle, it 
can impact precision electroweak measurements in unusual ways, due to its
chirality.    Significant bounds on  the interaction strength of
the $\nu_R$ can already be obtained
with presently available data.

      Since LEP II has set $M_{S}$ to be higher than the W mass at
 low energies we can integrate out the S
boson and perform a Fierz transformation to obtain the following 
effective four-fermion Lagrangian in terms of weak eigenstates:
\begin{eqnarray}
\label{eq:fourfermi}
-{\cal{L}}_{\mathrm S}^{eff} &=& \frac{1}{2M_{ S}^2} \bigg\{ \left| f_{12}
\right|^2\overline{ e_{L}}\gamma_{\rho}\nu_{e}
\overline{\nu_{\mu}}\gamma^{\rho}\mu_{L} + f_{12}f_{13}^{*}\overline{e_{L}}
\gamma_{\rho}\nu_{e}\overline{ \nu_{\tau}}\gamma^{\rho}\mu_{L} \\ \nonumber
& &- [f_{12}^{*}f_{23}\overline{e_{L}}\gamma_{\rho}\nu_{\tau}\overline{ 
\nu_{\mu}}\gamma^{\rho}\mu_{L} +f_{13}^{*}f_{23}\overline{e_{L}}
\gamma_{\rho}\nu_{\tau}\overline{\nu_{\tau}}\gamma^{\rho}\mu_{L}] \\ \nonumber
& &- g_1^*g_2\overline{e_{R}}\gamma_{\rho}\nu_{R}\overline{\nu_{R}}
\gamma^{\rho}\mu_{R} - g_1^*f_{23}\overline{e_{R}}\nu_{\tau L}
\overline{\nu_{R}}\mu_{L} \\ \nonumber
& &+ g_1^*f_{12}\overline{e_{R}}\nu_{eL}\overline{\nu_{R}}\mu_{L} - 
g_2f_{12}^*\overline{e_{L}}\nu_{R}\overline{\nu_{\mu}}\mu_{R} - g_2f_{13}^*
\overline{e_{L}}\nu_{R}\overline{\nu_{\tau}}\mu_{R} \\ \nonumber
& &+ 1/4[ g_1^*f_{23}\overline{e_{R}}\sigma_{\rho\lambda}\nu_{\tau}
\overline{\nu_{R}}\sigma^{\rho\lambda}\mu_{L} +g_2f_{13}^*\overline{e_{L}}
\sigma_{\rho\lambda}\nu_{R}\overline{\nu_{\tau}}\sigma^{\rho\lambda}\mu_{R} \\
\nonumber
& &+ g_2f_{12}^*\overline{e_{L}}\sigma_{\rho\lambda}\nu_{R}\overline{\nu_{\mu}}
\sigma^{\rho\lambda}\mu_{R} - g_1^*f_{12}\overline{e_{R}}\sigma_
{\rho\lambda}\nu_{e}\overline{\nu_{R}}\sigma^{\rho\lambda}\mu_{L}]\bigg\} 
+h.c. \;. 
\end{eqnarray}
We have shown only the terms relevant for $\mu$ decay. The dominant
term for the muon lifetime is then given by
\begin{equation}
\label{eq:SM}
-{\cal{L}}_{eff} = \frac{4G_{F}^{SM}}{\sqrt{2}}\left\{
g_{\mathrm LL}^{\mathrm V}\overline{e_{L}}\gamma_{\rho}\nu_{e\mathrm L}
\overline{\nu_{\mu\mathrm L}}\gamma^{\rho}\mu_{\mathrm L} + \ldots 
 \right\} \;,
\end{equation}
where
\begin{equation}
\label{eq:gll}
\begin{split}
g_{\mathrm LL}^{\mathrm V}& = 1 + \Delta g_{\mathrm LL} \\ 
& = 1 + \frac{2\left|f_{12}\right|^2 M_{W}^2}{g^2M_{S}^2} \:,
\end{split}
\end{equation}
and $g$ is the SU(2) gauge coupling.

We have adhered to the 
notations of the Particle Data Group \cite{Pdg}. For the purpose of this paper it
is convenient to work in the weak eigenbasis. 
One can easily see from 
Eq(\ref{eq:fourfermi}) that without 
$\nu_R$ the Zee model will give rise the same chiral structure as the SM [
see Eq.(\ref{eq:SM})]. The first term
will interfere coherently with the SM amplitude and is the most important
contribution to the muon lifetime, $\tau_\mu$. The terms involving $\nu_R$ 
add incoherently can be neglected in the muon lifetime $\tau_\mu$. 
Explicitly, \cite{Kin}\cite{ber}
\begin{equation}
\label{eq:taulife}
\frac{1}{\tau_\mu} =  \frac{G_F^{SM\,2}m_\mu^5}{192\pi^3}( 1 + 
\Delta g_{\mathrm LL})\left[ 1 + \frac{\alpha}{2\pi}(\frac{25}{4} - \pi ^2)\right] \;,
\end{equation}
where we have used the radiatively corrected expression and have neglected
terms involving $m_{e}^2/M_{W}^2$ and $m_{\mu}^2/M_{W}^2$. The SM fermi
constant is given by \cite{Sir}
\begin{equation}
\label{eq:GF}
G_{F}^{SM} =\frac{ \pi\alpha}{\sqrt{2}M_W^2\left( 1 - M_W^2 /M_Z^2 \right)
(1-\Delta r)} \;.
\end{equation}
Since $\tau_\mu$ is one of the most accurately determined quantity
in particle physics a careful analysis is need in order  
to obtain a bound on the coupling $f_{ij}$  versus M$_S$. In Eq. 
(\ref{eq:GF}) the fine
structure constant $\alpha(0)$ is very accurately known to be
1/137.03598959 and $\Delta r$ is the SM correction and is found to be \cite{Pdg2}
$\Delta r = .0349\mp.0019\pm.0007$. However, in computing $G_{f}^{SM}$ the largest error comes from $M_{W}$
measurement with current value given as $M_{W} = 80.39\pm.06$ GeV \cite{Kar}
and a much smaller error comes from 
$M_{Z} = 91.1867\pm.0020$ GeV. Although the error in W mass 
measurement is only 0.075\% it gets amplified in Eq.(\ref{eq:GF}) to 0.36\% and thus
constitutes the biggest uncertainty. In contrast, the error in $M_Z$
measurement gives an error of 0.015\% in $G_{F}^{SW}$.
From the above we can obtain the
error in the determination of $G_{F}^{SM}$ to be
\begin{equation}
G_{F}^{SM}= {(1.166\pm0.005)}\times10^{-5} \rm{GeV}^{-2}\;.
\end{equation}
In the above equations the LFV  physics
and the SM are assumed to be factorizable. 
This is a good approximation since the $S^-$ scalar
does not couple to the W, because it is an SU(2) singlet, and hence does not
affect the correction to the W propagator. We note that it
 does alter the running of $\alpha$, but the effect
is much smaller than the uncertainty in $M_W$ and can
be neglected. To get a bound on the couplings $f_{12}$ 
we demand that the corrections be no bigger the SM error
given above. We use Eq. (\ref{eq:gll}) to obtain a bound on 
$|f_{12}|^2$ as a function of $M_{S}^2$ which is displayed in Fig.(\ref{fig1}).

Besides $\tau_\mu$, the electron spectrum from muon decay also gives information
on possible new physics.  This is usually quoted in terms of the
Michel parameter measurements. It is interesting to note that only the 
following three Michel parameters get a correction from the Zee model
\begin{equation}
\begin{split}
1 - \frac{\xi\delta}{\rho} & \approx 2\left[ \left| g_{\mathrm RR}^{\mathrm V}\right|^2 + 
\left|g_{\mathrm LR}^{\mathrm S}\right|^2 \right] \\
&\geq 0   \; ,
\end{split}
\end{equation}
\begin{equation}
\begin{split}
1 - \xi^{\prime} & \approx 2\left[ \left| g_{\mathrm RL}^{\mathrm S}\right|^2 + 
\left|g_{\mathrm RR}^{\mathrm V}\right|^2 \right] \\
&\geq 0   \; ,
\end{split}
\end{equation}
and
\begin{equation}
\begin{split}
1 - \xi^{\prime\prime} & \approx 2\left[ \left|g_{\mathrm RL}^{\mathrm S}\right|^2 + 
\left|g_{\mathrm LR}^{\mathrm S}\right|^2 \right]\\
&\geq 0 \;,
\end{split}
\end{equation}
here
\beqa
\label{eq:grl}
\left|g_{\mathrm RL}^{\mathrm S}\right|^2 &=& \left|\frac{g_1M_W^2}{g^2M_S^2}\right|^2\left(\left|f_{12}\right|^2 
 + \left|f_{23}\right|^2\right) \leq 0.180 , \\
\left|g_{\mathrm LR}^{\mathrm S}\right|^2 &=& \left|\frac{g_2M_W^2}{g^2M_S^2}\right|^2\left(\left|f_{12}\right|^2
 + \left|f_{13}\right|^2\right) \leq 0.0156 , \\
\label{eq:grr}
\left|g_{\mathrm RR}^{\mathrm V}\right| &=& \frac{M_W^2\left|g_1g_2\right|}{g^2M_S^2} \leq 0.033  \; .
\eeqa
All the others take their canonical SM values. The left hand sides of Eq.(\ref{eq:grl} -
\ref{eq:grr}) are 
experimental bounds taken from \cite{Pdg}. One other noteworthy point
is that at the four-fermi level a right-handed current coupling is
induced because of the $S$ and $\nu_R$ interactions in lepton sector.  However,
this model does not have a corresponding coupling in the quark sector. 
This leads to a possible interesting  scenario in which an apparent 
contradiction can occur in the positive outcome of right-handed current searches in the purely
lepton sector such as
in  $\mu$ and $\tau$ decays and a negative outcome in similar searches using hadrons such as 
in $\beta$ decays of nuclei or pion decays. In
Fig. \ref{fig1} we also display the bounds on the 
couplings of the $S$ boson from Michel parameters. These bounds are complementary to 
the limit one obtains from the lifetime study.

Another tree level process that probes lepton number violation
is $\tau$ leptonic decay. The relevant effective four-fermi Lagrangian is
obtained from Eq.(\ref{eq:fourfermi}) by substituting the subscripts 2 to 3 and 1 to 3
appropriately. Here we use the branching ratio $R_{e\mu}$ defined by
$\tau \to e\nu\nu/\tau \to \mu\nu\nu$ and obtain 
[compare Eq.(\ref{eq:taulife})]
\begin{equation}
\label{eq:remu}
R_{e\mu} = 1 + 2 \frac{M_W^2\left(\left|f_{13}\right|^2 - \left|f_{23}\right|^2\right)}{g^2M_S^2} \;.
\end{equation}
Since experimentally $R_{e\mu}$ is close to unity the error 
$\pm 0.006$ \cite {Kar} only gives the absolute value
of the difference between $|f_{13}|^2$ and $|f_{23}|^2$. Notice that there is 
no dependence on the strength of the couplings involving $\nu_R$. Thus,
$\tau$ leptonic decays can be used to probe a
different region of parameter space from $\mu$ decays. Moreover, 
Eq.(\ref{eq:remu}) shows that the $f_{13}$ and $f_{23}$ are
of the same order of magnitude. We also find the Michel parameter 
measurements here
place less stringent bounds than from $\mu$ decays.

Next we discuss the one loop effect involving the S-scalar. The
most accurately measured LFV neutral current process
is the decay $\mu\to e\gamma$. The branching
ratio is calculated to be
\begin{equation}
Br( \mu\to e\gamma) = \frac{\alpha\left(\left|g_1g_2\right|^2 + \left|f_{13}f_{23}\right|^2\right)}
{24\pi g^4}\left(\frac{M_W}{M_S}\right)^4 \;,
\end{equation}
This is sensitive to $|g_{1}g_{2}|^2 +|f_{23}f_{13}|^2$ and provides
the most stringent limit \cite{Bol} on these parameters; see Fig. \ref{fig2}. 
Since this 
is a sum of two positive terms and implies that $|g_1g_2|$ and $|f_{13}f_{23}|$
must be of order 0.01 for $M_S = 800$ GeV. We note in passing that similar $\tau$
decays  will test different combinations of coupling constants.

A similar calculation gives the correction to the anomalous
magnetic moment of the charged leptons, $a_\ell$ where $\ell = e,\mu,$ or $\tau $. Then
\begin{equation}
\label{eq:magmom}
a_{\ell} = \frac{\sum_{\j \neq \ell}\left|f_{\ell \mathrm j}\right|^2 + \left|g_{\ell}\right|^2}{96\pi^2}
\left(\frac{m_\ell}{M_S}\right)^2 \;.
\end{equation}
Both $a_e$ and $a_\mu$ are very well measured quantities. However, since scalar 
interactions
flip chirality and because of dimensional consideration we obtain two powers  
of lepton mass in Eq.(\ref{eq:magmom}).  These powers of lepton mass, when
combined with the loop suppression factors make 
these limits less constraining than some of the other processes considered here.
Some of the other processes considered here, such as the $\mu$ decay 
parameters and the right and left handed lepton coupling to the Z 
discussed below, set tight limits on $|f_{12}|^2$, $|f_{13}|^2$, and 
$|f_{23}|^2$. Therefore, 
 we can use Eq (\ref{eq:magmom}) to put limits on $g_{1}$ and $g_{2}$. The
limits from measurements of the electron and muon are shown in Fig. \ref{fig3}. If the error
can be reduced by a factor of 20 on $a_{\mu}$ the limit obtained is
the lower curve in Fig. \ref{fig3}. This is the target precision  of the Brookhaven
experiment E821 \cite{E821}.

We proceed to discuss the effect of the Zee model on precision measurement
at the Z-pole. The main effect is on the left-right asymmetry of
the charged leptons and lepton flavor changing neutral current decays.   
To one loop order 
the correction to the $Z\ell\overline \ell$
vertex, $\delta\Gamma_{\mu}$ is given by
\begin{equation}
\delta\Gamma_{\mu} = \delta g_{\mathrm L}\gamma_{\mu\mathrm L} + \delta g_{\mathrm R}\gamma_{\mu\mathrm R}\;,
\end{equation}
where
\begin{equation}
\label{eq:deltagl}
\delta g_{\mathrm L} = \frac{b}{96\pi^2}\left(\sum_{j \neq \ell}\left|f_{\ell \mathrm j}\right|^2
\right)\left(\frac{1}{3}s_W^2 - \frac{1}{3} - \ln b + i\pi\right) \;,
\end{equation}
and
\begin{equation}
\label{eq:deltagr}
\delta g_{\mathrm R} = - \frac{g_{\ell}^2bs_W^2}{288\pi^2} \;.
\end{equation}
In the above we have used $b=M_Z^2/M_S^2$, $s_W = \sin\theta_W$ and $c_W = \cos\theta_W$. As expected the
 left-handed coupling is
corrected in the Zee model without $\nu_{R}$. However, if $\nu_{R}$
exists it will also modify the right-handed electron coupling to
the Z in spite of the fact that, although $\nu_{R}$ is light, there is no 
tree level 
$Z\nu_{R}^{c}\nu_{R}$ coupling. Using 
Eqs.(\ref{eq:deltagl}) and (\ref{eq:deltagr})
one can calculate the left-right asymmetry, $A_{\mathrm LR}$. It is given by
\begin{equation}
A_{\mathrm LR}= A_{\mathrm LR}^0\bigg\{ 1 +\frac{ 32s_W^2(1-2s_W^2)[2s_W^2\delta g_{\mathrm L}-(1
-2s_W^2)\delta g_{\mathrm R}]}{(1 - 4s_W^2)[1 +(1-4s_W^2)^2]} \bigg\}     \;,
\end{equation}
where $A_{\mathrm LR}^0$ is the SM left-right asymmetry. 
The most stringent bounds are
obtained by combined LEP \cite {Kar}and SLC measurements 
\cite {Abe} given in
terms of the axial and vector couplings. This is shown in Fig. \ref{fig4}.

With the above limits on the various couplings we can also
predict the upper limits for the rare decays $Z \to \overline{l}_{i}
l_{j}$ where $l_{i}\neq l_{j}$ and $l_{i} = e,\mu, \tau$. The width is
calculated to be 
\begin{equation}
\Gamma (Z \to \overline {l}_i l_j) =\frac{g^2 M_Zb^2}{24\pi^5(288c_W)^2}\bigg\{\left|f_{ik}f_{jk}\right|^2\left[(3 \ln b -
c_W^2)^2 + \pi^2\right] +\left|g_ig_j\right|^2s_W^4 \bigg\} \;,
\end{equation}
where $i\neq j \neq k$.
With the limits given by the above considerations the 
upper limits for the $ e\mu $, $\mu\tau$, and $e\tau $ are all unmeasurably
small in the foreseeable future. For example the upper limit for the Z $\to e \mu$ branching ratio is of order
$\mathrm 10^{-10}$ for
a 800 GeV bilepton scalar. 

Having examined the constraints on the bilepton scalar we proceed to
look into future tests. Short of discovering the $S$ scalar itself, which is
best done in a linear $e^+e^-$ collider, we
propose that e, $\mu$ ,$\tau$ universality for the W boson will be a good
place to look this kind of LFV physics. Explicitly,
the branching ratios $W \to e\nu / W \to \mu\nu$ and $W \to e\nu /W \to
\tau\nu$ will be very valuable. In the SM these branching ratios are
unity at the tree level but at the level of first order radiative correction
 this universality is broken.
However, this breaking is suppressed by the factor of \cite{App} 
$\alpha(m_{\mu}^{2}/M_{W}^2)$ or $\alpha(m_{\tau}^2/M_{W}^2)$ and hence
very small. The important point here is that the radiative corrections 
are accurately predicted in the SM, and the measurements of these branching ratios
are very clean theoretical probes of LFV physics. We illustrate this in the Zee model 
by calculating the correction to the W leptonic decay widths. We find
\begin{equation}
\label{eq:wdecay}
\begin{split}
 \Gamma(W\to l_{i}\nu_{i}) =& \frac{g^2M_W}{48\pi}\left[1 - \frac{\alpha}{2\pi}\left(\frac{2\pi^2}{3}-\frac{77}{12}\right)
\right]  \\
& \bigg\{ 1+ \frac{M_W^2}{72\pi^2M_S^2}\left(1-3\ln\frac{M_W^2}{M_S^2}\right)\left[\sum_{j\neq i}
\left|f_{ij}\right|^2\right]\bigg\} \;,
\end{split}
\end{equation}
where $i=e, \mu$ ,or $\tau$. The first line is the radiatively corrected W boson
leptonic width \cite{Marc} and the second term in the curly bracket is the 
Zee model correction. 
As expected the
left-handed nature of the coupling is preserved and $\nu_R$ does not play a role. We
have also kept only the leading term in $M_W^2/M_S^2$ in the calculation and neglected
the lepton masses.
Eq. (\ref{eq:wdecay}) immediately gives rise to the following results for the leptonic
branching ratios:
\beqa
\label{eq:wbr}
\mathrm Br \left(\frac{W \to \mu \nu}{W \to e \nu}\right) = 1 + k(\left|f_{23}\right|^2 -\left|f_{13}\right|^2) , \\
\mathrm Br \left(\frac{W \to \tau \nu}{W \to e \nu}\right) = 1 + k(\left|f_{23}\right|^2 -\left|f_{12}\right|^2) , \\
\mathrm Br \left(\frac{W \to \tau \nu}{W \to \mu \nu}\right) = 1 + k(\left|f_{13}\right|^2 -\left|f_{12}\right|^2) \; ,
\eeqa
where $k= \frac{M_W^2}{72\pi^2M_S^2}\left(1+3\ln\frac{M_S^2}{M_W^2}\right)$. As seen from $\tau$ universality
the combination $\left|f_{23}\right|^2 - \left|f_{12}\right|^2$ is bounded to be small and so we
determine that the first branching ratio of Eq. (\ref{eq:wbr})
can only accommodate  a correction of less than 2$\times \mathrm 10^{-5}$ for $M_S$= 800 GeV. Furthermore
the constraints we obtained on $\left|f_{13}\right|^2$ and $\left|f_{23}\right|^2$ from the Z pole allow
the universality violation in the last two branching ratios of Eq. (\ref{eq:wbr}) to be as large as 1$\times
\mathrm 10^{-3}$. 
At the LHC and the NLC where large samples of W decays are expected, these
decays will be the most important and cleanest tool for probing LFV in the
charged current sector.  We note that the
uncertainty due to the t-quark that plagues $\Delta r$ in the interpretation
of the muon lifetime measurement does not enter here at the one loop level.  
Once sufficient statistics are obtained these measurements 
will supersede many low energy tests of LFV.

        The last stop in our discussion of futuristic experiments is the production of the $S$-scalar.
 In particular we
focus on pair production in $e^+e^-$ colliders. This is very similar to charged Higgs production in
the two doublets model \cite{Kom}. Since the Yukawa couplings $f_{12}$ and $f_{13}$ are seen to be
small the t-channel process involving neutrino exchanges can be neglected and one needs only to consider
the s-channel virtual photon and Z exchange graphs. The coupling are all determined by the SM 
charges [see Table I] the production cross section can be calculated with $M_{S}$ 
as the only free parameter.
Explicitly, the cross section for $e^+e^- \to S^+ S^-$ is 
\begin{equation}
\sigma = \frac{\pi \alpha ^2\beta^3}{3s}\left[ 1 +\frac{s(-1 + 4s_W^2)}{2c_W^2(s - M_Z^2)} +
\frac{s^2(-1+4s_W^2 + 8s_W^4)}{4c_W^4(s - M_Z^2)^2}\right] \; ,
\end{equation}
where $s$ is the cm energy squared and $\beta =\sqrt { 1 - 4M_{S}^2/s }$. If the mixing of the
bilepton scalar with charged Higgs bosons can be
neglected its dominant decay mode would be into $l_i \nu_j$. The signal will be $\tau$ and $\mu$ plus missing energy 
and unmistakable.

        As seen from the results we have presented, a bilepton scalar induces many interesting and testable
effects. We have employed the Zee model augmented with a light $\nu_{\mathrm R}$ for quantitative studies since
it is relatively simple and the number of free parameters are relatively few. We find that the strongest bound on 
the Yukawa couplings in the model comes from $\mu \to e \gamma$ decay, lepton universality in $\tau$ decays and the lifetime of the muon. In
particular $f_{12}$ and $f_{13}$ are both constrained to be less than 
of order 0.1 and 1 respectively 
for $M_S=$800 GeV. 
However, the couplings
$g_i$ have much looser bounds. Since Yukawa couplings are in general not universal we expect leptonic 
universality to be violated in W decays (see Eq. (\ref{eq:wdecay})). This illustrates the importance of W decays in
probing LFV physics.
 Another effect which will be harder to discern 
is the modification
 of the U(1) coupling constant running. The $S$-scalar contributes an amount of 
$\frac{g^{\prime 3}}{48\pi^2}$ to the 
U(1) $\beta$ function where $g^{\prime}$ is the U(1) gauge coupling. This will upset the unification of the SM gauge couplings
at very high energies  which is usually taken to be 
a hint for supersymmetry. We do not consider this to be a serious impediment for the following reason. Although the 
Zee model is interesting in its own right, it carries the same arbitrariness as the Higgs
sector of the SM. Hence, we expect it be part of a larger structure that entails supersymmetry. Promoting
the bilepton scalar to a superfield necessitates the introduction of a second scalar in order to cancel the
anomaly caused by the fermionic partner of the $S$. In short one would have to enlarge the minimal supersymmetric
standard model by two bilepton superfields and also one singlet neutrino superfield. The details of how
this can be achieved and the ensuing intricate phenomenology is beyond the scope
of the present paper and
 we shall defer the study of this issue to a later
work.

This work has been supported by the Natural Sciences
and Engineering Research Council of Canada.
\newpage
\begin{table}
\caption{ Lepton number, SU(2), and U(1) charges of the left-handed lepton doublet, the $\nu_R$, the charged lepton
 conjugate, and the S-scalar}
\renewcommand{\arraystretch}1
\begin{center}
\begin{tabular*}{100mm}{@{\extracolsep{\fill}}|c| c c c c|}   \hline
&$\ell$
&$\nu_R$
&$\ell ^c$
&$ S^-$
\\
\hline
SU(2) &$1/2$ &0 &0 &0\\
\hline
Y & -1 & 0 &-2 & -2 \\
\hline
L & 1 & 1 & 2  & 2 \\
\hline
\end{tabular*}
\end{center}
\label{tab1}
\end{table}

\clearpage

\baselineskip 16pt plus 2pt minus 2pt

\begin{figure}
\epsfxsize=14cm
\epsfxsize=14cm
\caption{Bounds on coupling constants derived from measurements of
Michel parameters for 
muon decay are plotted as a function of the mass of the scalar, $M_S$.  
The constraint on $|f_{12}|^2$ comes from the measurement of the
muon lifetime.}
\label{fig1}
\end{figure}

\begin{figure}
\epsfxsize=14cm
\epsfxsize=14cm
\vskip 0.5in
\caption{As in Figure 1, for the constants 
$|f_{23} f_{13}|^2 + |g_{1} g_{2}|^2$.  This constraint comes from 
the experimental limit on the width of the 
decay process $\mu \rightarrow e \gamma$.    
}
\label{fig2}
\end{figure}

\begin{figure}
\epsfxsize=14cm
\epsfxsize=14cm
\vskip 0.5in
\caption{Coupling constants plotted as in Figure 1.  These 
constraints come from precision measurements of the 
anomalous magnetic moment
of the electron (top line), and of the muon (middle line).  The lower line
shows the bound which could be obtained if the muon measurement were 
twenty times more precise.    
}
\label{fig3}
\end{figure}

\begin{figure}
\epsfxsize=14cm
\epsfxsize=14cm
\vskip 0.5in
\caption{Coupling constants plotted as in Figure 1.  This plot represents
three different constraints as determined by the choice of leptons,
$e, \mu, \tau$ for the parameters, $l,j,k$ (with $l \neq j \neq k$).
These constraints
are derived from the measurements of leptonic 
axial and vector couplings.    
}
\label{fig4}
\end{figure}

\epsfxsize=14cm
\epsfxsize=14cm
\centerline{\epsffile{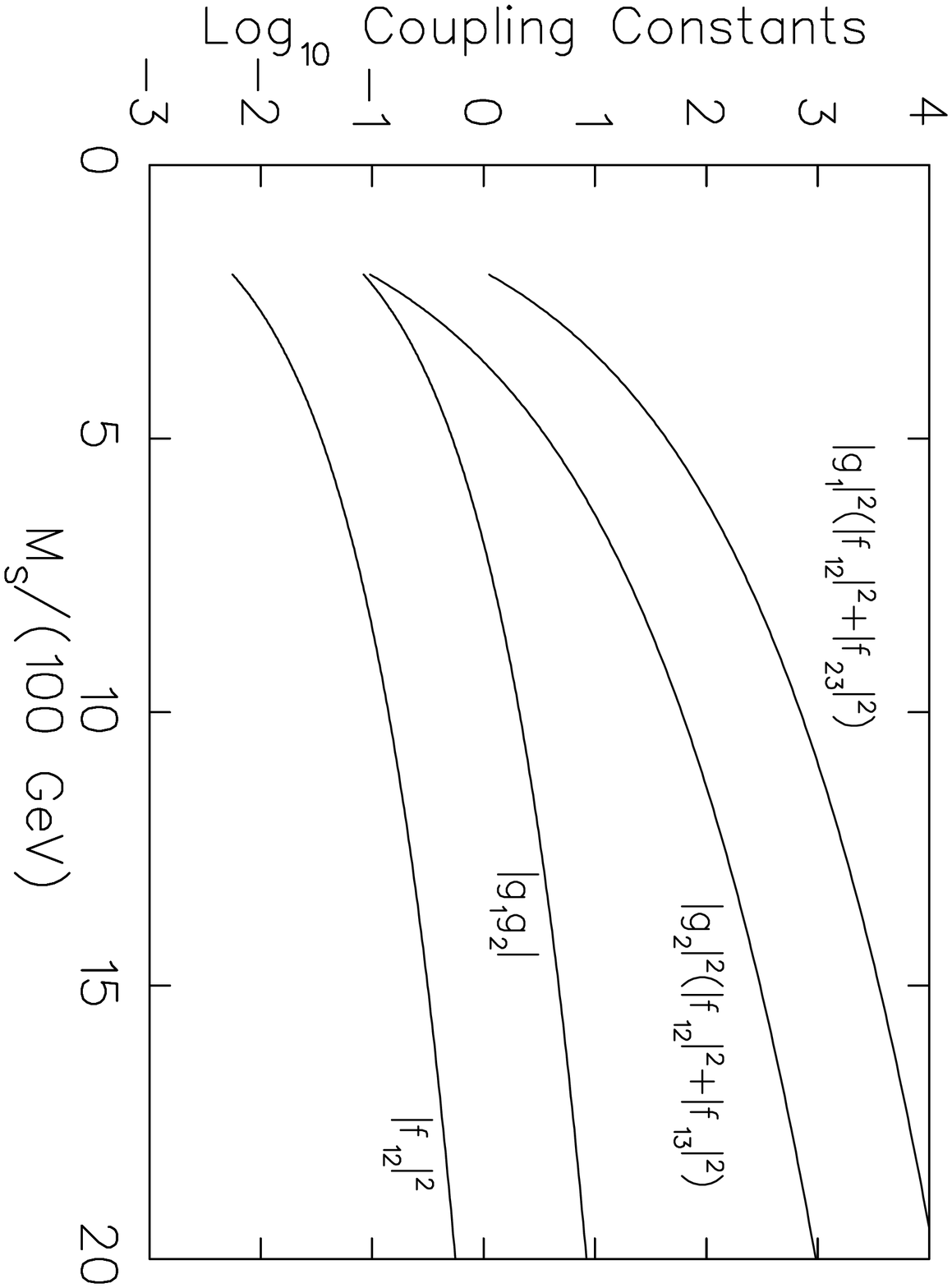}} 
\epsfxsize=14cm
\epsfxsize=14cm
\centerline{\epsffile{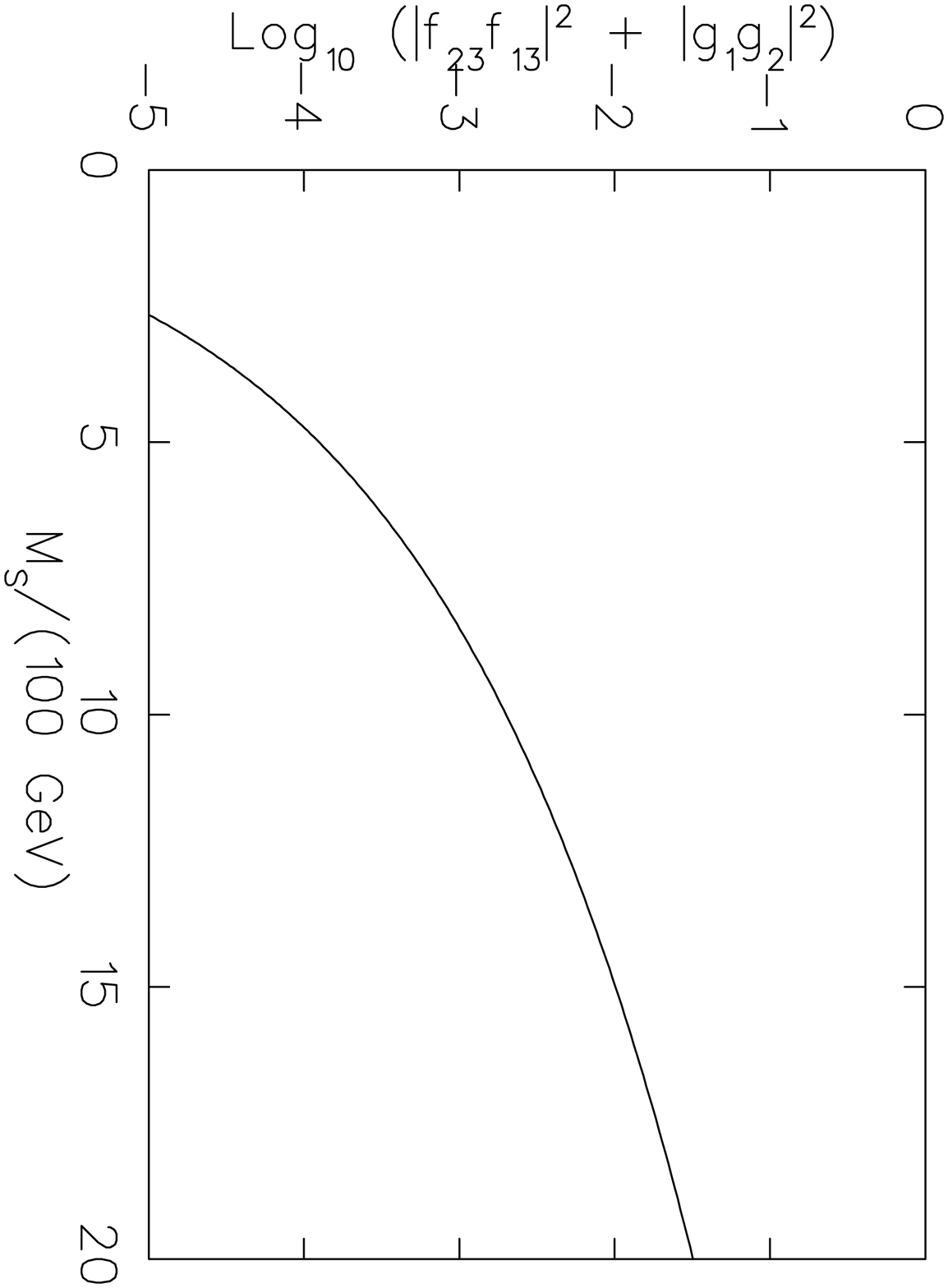}}
\epsfxsize=14cm
\epsfxsize=14cm
\centerline{\epsffile{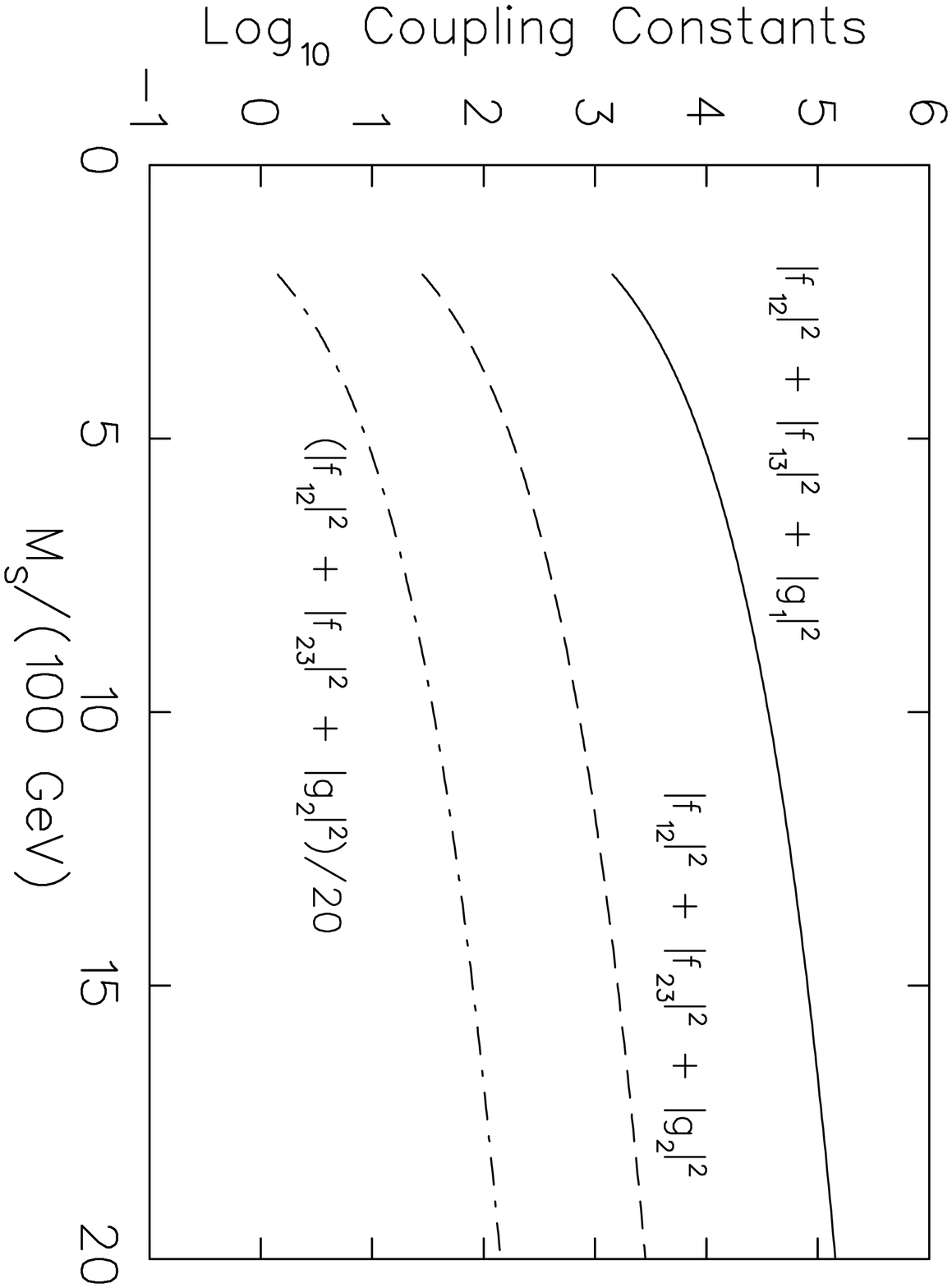}}
\epsfxsize=14cm
\epsfxsize=14cm
\centerline{\epsffile{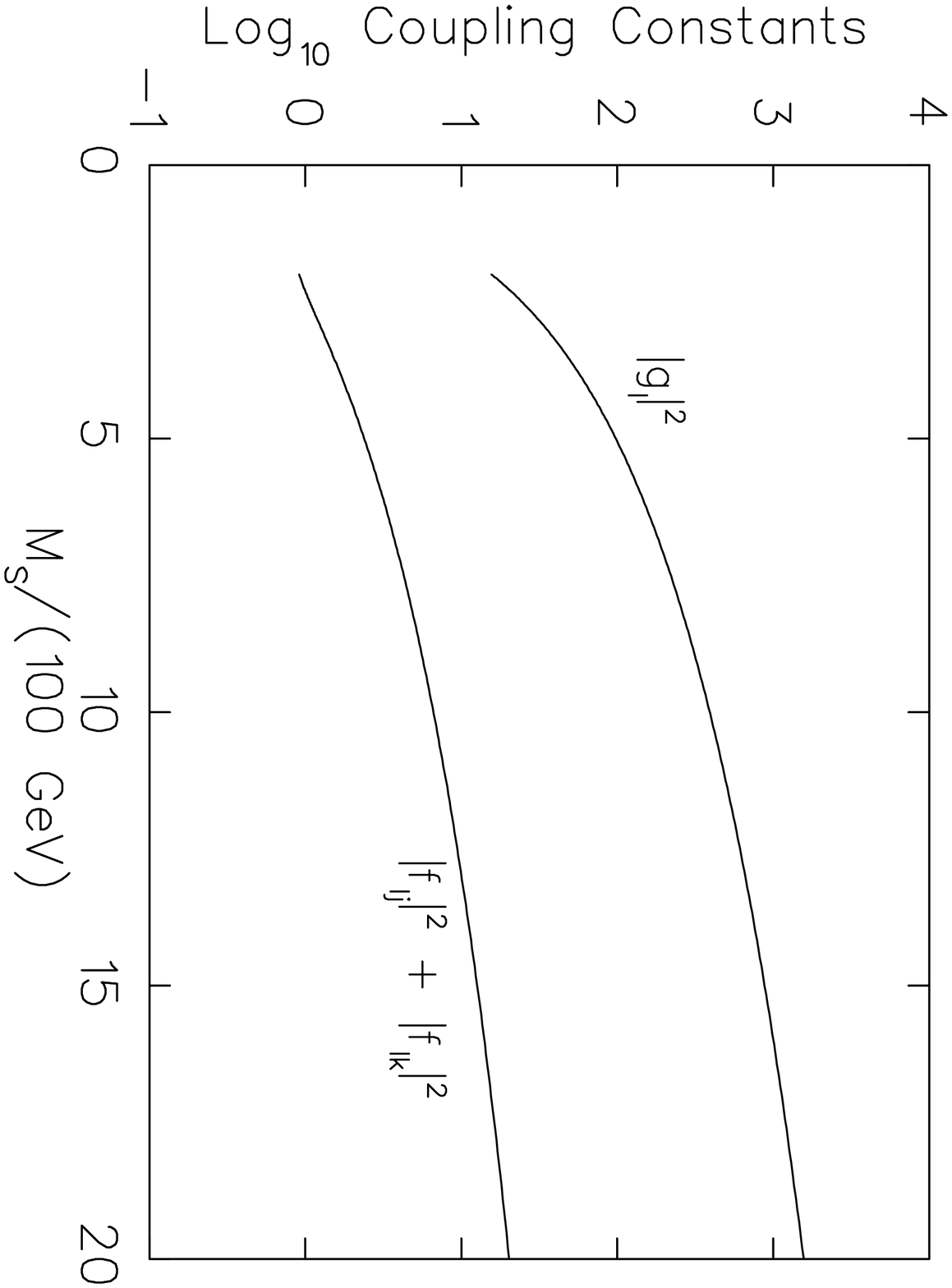}}


\begin{thebibliography}{99}
\bibitem{Con}For a review see J.M. Conrad Proc.29th ICHEP Conf.,
             Vancouver, Canada, July (1998), hepex 9811009
\bibitem{Naka}M. Nakahata. Talk at Workshop on Nuclear Reactions in the stars and in the laboratory,
              Feb. 1999, Trento, Italy
\bibitem{Jarl}C. Jarlskog, Phys. Lett 241B (1990) 579
\bibitem{Chang}L.N. Chang, D.Ng, and J.N. Ng, Phys. Rev. D50 (1994) 4589
\bibitem{Zee}A. Zee, Phys. Lett. 93B, (1980), 389; 161B (1985) 141.
\bibitem{Wolf}W. Wolfenstein, Nucl. Phys. B173 (1980) 93
\bibitem{Smir}A. Yu. Smirnov and M. Tanimoto, Phys. Rev. D55 (1997) 1665.
\bibitem{Jarl2} C. Jarlskog, M. Matsuda, S. Skadhauge, and M. Tanimoto,
               hepph 9812282 (1998).
\bibitem{fugu}M. Fukugita and T. Yanagida, Phys. Rev Lett. 58 (1987) 1807
\bibitem{E614}D. Gill, Triumf Experiment E614
\bibitem{Pdg}Particle Data Group, Euro. Phys. 3C (1998) 282
\bibitem{Kin}T. Kinoshita and A. Sirlin, Phys. Rev. 113 (1959) 1652
\bibitem{ber}S. Berman, Phys. Rev. 112 (1958) 267
\bibitem{Sir}A. Sirlin, Phys. Rev. 29D, (1984) 89
\bibitem{Pdg2} Particle Data Group, Euro Phys. 3C (1998) 91
\bibitem{Kar}D. Karlin, Proc. 29th ICHEP Conf., Vancouver Canada July (1998) 
\bibitem{Bol}R.D. Bolton et al, Phys Rev D38 (1988) 2077
\bibitem{E821} B.L. Roberts et al, Proc. XXVIII Int Conf on High Energy
               Physics, Warsaw, Poland, (1996). World Scientific,(1997). 1035
\bibitem{Abe}K. Abe et al, Phys. Rev. Lett 78 (1997) 2075
\bibitem{App}T. Appelquist,J.R. Primack, and H.R. Quinn, Phys. Rev. 8D 
             (1973) 2997.
\bibitem{Marc} W.Marciano, Phys. Rev. 12D (1975) 3861
\bibitem{Kom}S. Komaniya, Phys Rev D38 (1988) 2158
\end{thebibliography}
\end{document}